\documentclass[]{spie}  

\usepackage{amsmath,amsfonts,amssymb}
\usepackage{graphicx}
\usepackage[colorlinks=true, allcolors=blue]{hyperref}
\usepackage{wrapfig}
\usepackage[caption=false]{subfig}

\title{Impact of local turbulence on high-order adaptive optics}

\author[1]{Hugo Nowacki}
\author[1]{Jean-Baptiste Le Bouquin}
\author[2]{Carole Gouvret}
\author[2]{Aurélie Marcotto}
\author[2]{Sylvie Robbe-Dubois}
\author[1]{Karine Perraut}
\author[1]{Yves Magnard}
\author[1]{Alain Delboulbé}
\author[1]{Eric Stadler}
\author[1]{Sylvain Guieu}
\author[1]{Sylvain Rochat}
\author[1]{Didier Maurel}
\affil[1]{Univ. Grenoble Alpes, CNRS, IPAG, 38000 Grenoble, France}
\affil[2]{Université Côte d'Azur, Observatoire de la Côte d'Azur, CNRS, Laboratoire Lagrange, France}

\authorinfo{Further author information: (Send correspondence to Hugo Nowacki)\\
E-mail: hugo.nowacki@univ-grenoble-alpes.fr \\ Mailing adress : Hugo NOWACKI, IPAG, 414 rue de la piscine, 38400 Saint-Martin-d'Hères, FRANCE.}

\begin{document} 
\maketitle

\begin{abstract}
We present an experiment set to address a standard specification aiming at avoiding local turbulence inside the Coudé train of telescopes. Namely, every optical surface should be kept within a 1.5$^\circ$ range around ambient temperature. Such a specification represents an important concern and constraint when developing optical systems for astronomy. Our aim was to test its criticality in the context of the development of the VLTI/NAOMI and VLTI/GRAVITY+ adaptive optics. This experiment has been conducted using the hardware of the future Corrective Optics (CO) of GRAVITY+. Optical measurements were performed in order to observe the evolution of turbulence in front of a flat mirror for which the surface temperature was controlled in a range of $22^\circ$ above ambient temperature. A time-dependent analysis of the turbulence was led along with a spatial analysis. This experiment shows no influence of temperature on local turbulence. It should be noted however that this result is only applicable to the very specific geometry described in this paper, which is representative of an adaptive optics (AO) system located inside the Coudé train (facing-down mirror heated on its backface).
\end{abstract}


\section{INTRODUCTION}
\label{sec:intro}  

A standard specification to avoid local turbulence is to keep all optical surface within $\pm$ 1.5$^\circ$ with respect to the ambient temperature. This specification can be challenging for active devices. This is especially true for Deformable Mirrors based on the voice-coil technology, whose actuators produce a significant amount of heat very close to the optical surface. This technology is currently used for instance for the Adaptive Optics of the CHARA \cite{2020SPIE11446E..22A}, the VLTI Auxiliary Telescope arrays \cite{2020SPIE11446E..06H,2019A&A...629A..41W}, several secondary mirrors, and is the baseline for the GRAVITY+ project\footnote{https://www.mpe.mpg.de/7480772/GRAVITYplus\_WhitePaper.pdf}. 

This specification can also be challenging for optical devices preferentially operated at typical laboratory temperatures ($20^\circ$C), but which are deployed inside the Coudé train of telescopes where the ambient air temperature ranges from $0^\circ$C to $20^\circ$C. For a Deformable Mirror in which the stiffness is linked to the temperature, such a requirement imposes to provision larger margins on the critical trade-off between speed, stroke and heat. Yet, the NAOMI project for the VLTI-AT array ultimately had to implement a system to maintain the Deformable Mirror at a constant temperature of about $20^\circ$C, in order to achieve the desired stroke. This heating system deviates from the initial requirement of $\pm1.5^\circ$C for the surface temperature, without noticeable loss in performance. It questions the adequacy of this tight specification for the specific case of a Deformable Mirror which is ultimately part of an AO system correcting the turbulence.

\newpage
\section{INSTRUMENTAL SET-UP}
\label{sec:setup} 

\begin{figure} [ht]
   \begin{center}
   \begin{tabular}{c} 
   \includegraphics[width=\linewidth]{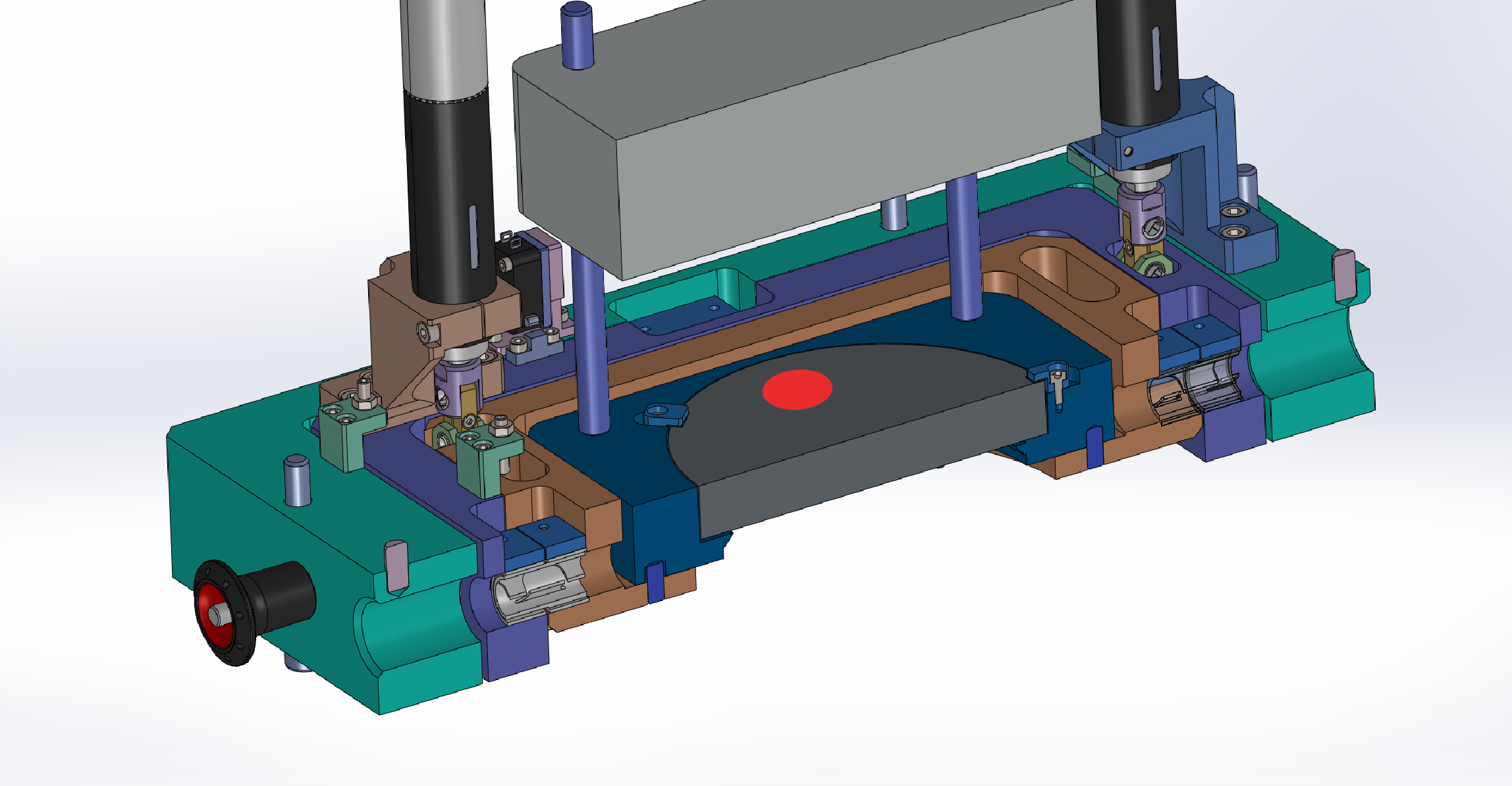}
	\end{tabular}
	\end{center}
   \caption[\label{fig:monture}]
   {Sliced view of the gimbal without the Corrective Optics Mount. The red circle shows the position of the heater.}
   \label{fig:monture}
   \end{figure} 
This section describes the hardware and the software of the experiment.

The experiment was executed in the actual test bench of the GRAVITY+ project, at the Observatoire de la Cote d'Azur \cite{2022_GRAVITYPLUS_BENCH}. The bench features several internal lights, and a long beam focalisation to create a pupil of about 100mm in the Deformable Mirror with a f-number of about 50. The Deformable Mirror is installed in a mount called the Corrective Optics (CO), where it is facing down. This CO is the original model that will be used inside the Coudé train of the Unit Telescopes (UTs) of the Very Large Telescope Interferometer (VLTI). The software part is a combination of both the software used for general testing of the CO (beyond this experiment) and dedicated software used/developed for this experiment.

\subsection{Hardware}
This mount can be decomposed into the mounting's structure, called the Corrective Optics Mount (COM) and the gimbal inside of it. The gimbal (see Fig\ref{fig:monture}) is the piece in which the Deformable Mirror will be inserted. It can be controlled to compensate for tip-tilt misalignment.

The final device of GRAVITY+ will be a 41x41 Deformable Mirror from ALPAO with a pitch 2.5mm for an active pupil of 100mm in diameter, and a clear aperture of 106mm. For the purpose of this experiment, we did not use a Deformable Mirror but a flat thick mirror instead (half grey cylinder on Fig.\ref{fig:monture}). Attached to this flat mirror is a mass placed in height so that it reproduces the geometry and mechanical constraints of the actual Deformable Mirror while leaving the mirror's backface accessible.

In the final configuration, the heat will be generated by the actuators of the Deformable Mirror, especially in bad seeing conditions. In the case of the flat mirror, we had to implement an external source of heat in order to be able to drive the temperature of the optical surface. We used a 40W heater\footnote{Reference : KHLVA-202/10 FLEXIBLE HEATER, POLYMIDE, 40W, 28VAC} glued to the backface of the flat mirror (this backface is thus facing up). Then, a Pt100 probe was placed on the same side of the mirror as the heater, at a distance comparable to the mirror's thickness so the monitored temperature corresponds to the optical surface's. A second Pt100 probe was used to check the ambient temperature. Overall, the whole set-up is to mimic the behaviour of a Deformable Mirror implemented in a realistic facing-down configuration (actually the exact one of GRAVITY+) and generating its own source of heat.

It should also be noted that the gimbal has a $13.26^\circ$ inclination with respect to the horizontal. We fixed the heater in the lowest part of the mirror, but we do not expect this to be important because the overall mirror should have a rather uniform temperature. The expected behaviour is that the heated surface would produce a flow of hot air in front of the field of view. This moving bubble would hence produce local turbulence. To observe this turbulence itself, we used a SID4\textsuperscript{\textregistered} wave front sensor \footnote{SID4 version 4.3 produced by PHASICS \copyright} that produce real-time intensity and phase map at a 7.5 Hz rate on a sensor of 3.6 x 4.8 mm$^2$ sampled with 120 x 140 pixels$^2$. This device was placed in the nominal optical path as designed for GRAVITY+. At this place, the pupil (106mm clear aperture) is slightly larger than the PHASICS wavefront sensor. We thus restrict the analysis to a reduced pupil of about 90mm in diameter.

\subsection{Software}
We monitored both the mirror's temperature and the power delivered to the heater thanks to a Python routine sent through the Beckhoff\textsuperscript{\textregistered} controller of the gimbal mount. We recovered phase maps thanks to the software of the SID4\textsuperscript{\textregistered} wave front sensor. A description of the maps we obtained is provided in section \ref{sec:measurements}.

\section{MEASUREMENTS}
\label{sec:measurements} 
We performed a series of measurements covering the range of $+0^\circ$ to $+22^\circ$ above ambient temperature with a $2^\circ$ step between each measurement. Every temperature is tested twice. A first time during the increase in temperature of the mirror, and a second time during the decrease in temperature of the mirror, which is not driven, but only happens in a passive way. The warm up sequence took about 3h, and the cool down sequence about 4h. Altogether, this gives a set of 23 observations that are analysed in section \ref{sec:analysis}.

Each observation consists into the acquisition of a cube of 512 continuous phase images taken with the SID4\textsuperscript{\textregistered} at a 7.5 Hz rate. We obtain a 68.27 s film showing the evolution of local turbulence in front of the mirror. This is trade-off between the temporal resolution, resolution in frequency, and the size of the individual files.

For illustration, we extract a series of snapshots in Fig.\ref{fig:snaps} where we have subtracted the average image to each snapshot. These snapshots were taken at a fifth of the nominal rate (1.5 Hz) to make the speed and amplitude of the changes we observe visible. Time goes from the left to the right and from up to bottom. One can see the appearance of a bubble on the right of the field which is then transferred to the left. The process takes the 6 images to happen hence a timescale of $\sim 4$ seconds for the turbulence as a rough first estimate (see section \ref{sec:timeanalysis} for more precise analysis of the time-dependant behaviour of the turbulence). The overall wavefront residual in the images is about 20\,nm, which is smaller than the typical fitting error budget in modern high resolution adaptive optics. Our experiment should therefore be able to detect any additional problematic turbulence that may arise from the heating of the mirror.

The following analyses of the turbulence in section \ref{sec:analysis} below are entirely based on these measurements.

\begin{figure} [ht]
   \begin{center}
   \begin{tabular}{c} 
   
   \includegraphics[width=0.33\linewidth, trim=52 0 59 0,clip]{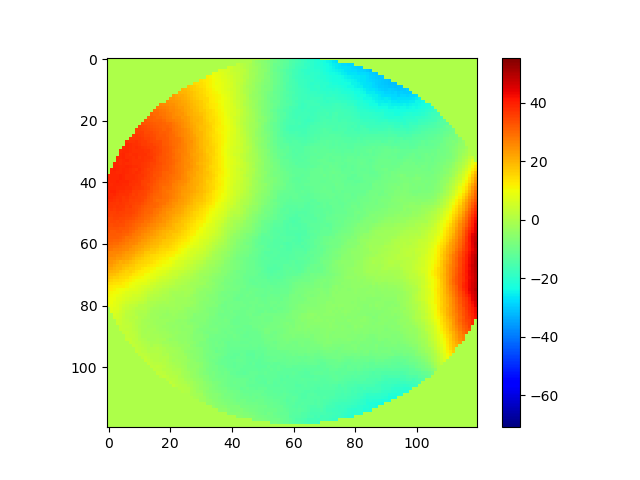}
   \includegraphics[width=0.33\linewidth, trim=52 0 59 0,clip]{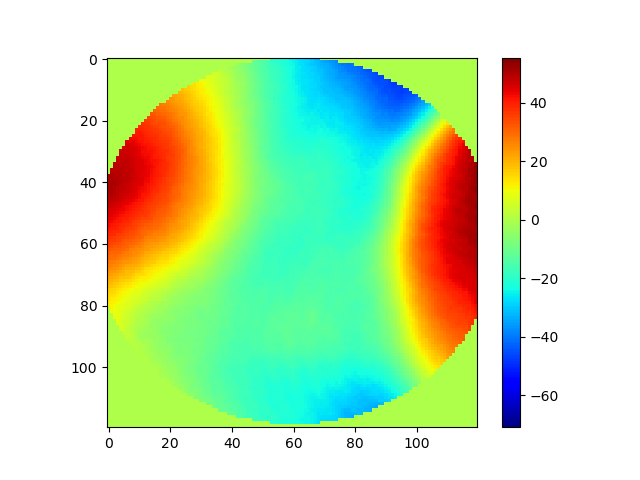}
   \includegraphics[width=0.33\linewidth, trim=52 0 59 0,clip]{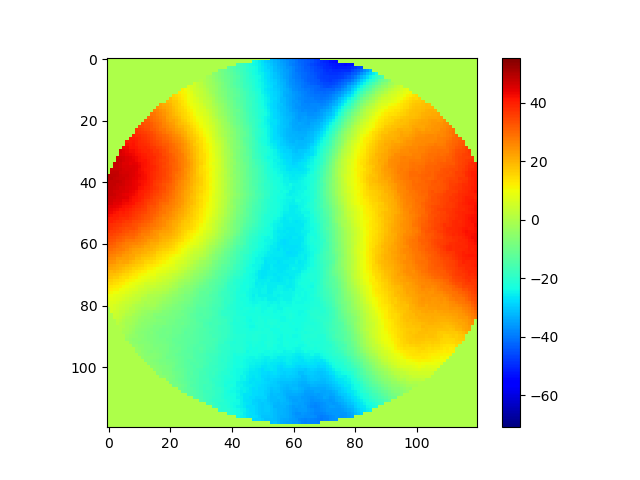}
   \end{tabular}
   \end{center}
   \end{figure}

\begin{figure} [ht]
   \begin{center}
   \begin{tabular}{c} 
   \includegraphics[width=0.33\linewidth, trim=52 0 59 0,clip]{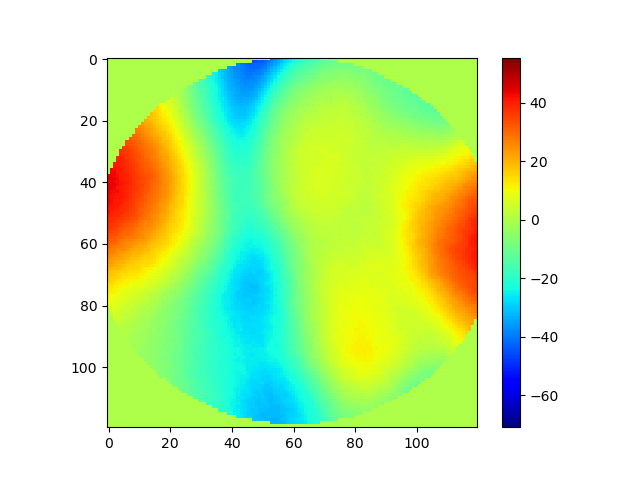}
   \includegraphics[width=0.33\linewidth, trim=52 0 59 0,clip]{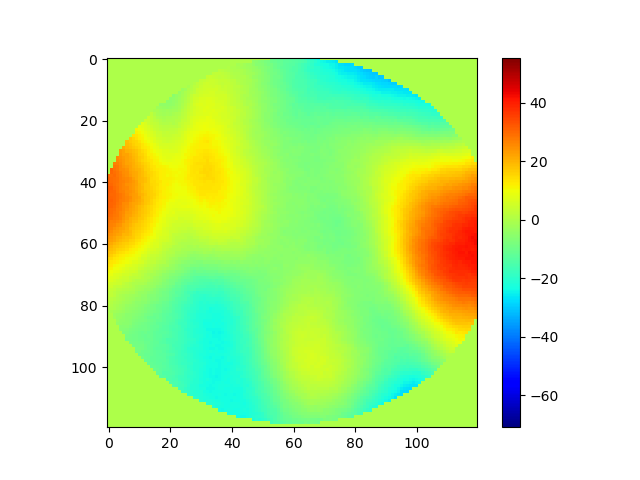}
   \includegraphics[width=0.33\linewidth, trim=52 0 59 0,clip]{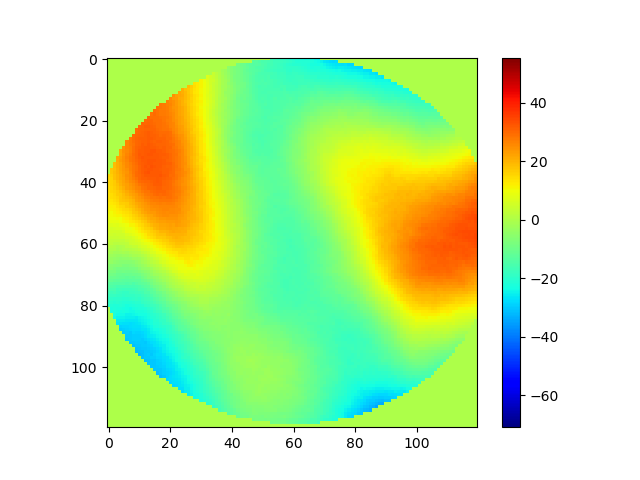}
	\end{tabular}
	\end{center}
   \caption[\label{fig:snaps}]
   {Six snapshots taken at 1.5 Hz to illustrate the turbulence at ambient temperature (the entire cube is recorded at a frame rate of 7.5 Hz). Time goes from the left to the right and from up to bottom. Color scale is constant from one picture to the next and gives the path difference in nm.}
   \label{fig:snaps}
   \end{figure}

\section{RESULTS/ANALYSIS}
\label{sec:analysis}

\subsection{Time-dependant analysis}
\label{sec:timeanalysis}
\begin{figure}[ht]
    \centering
    \includegraphics[width=8cm]{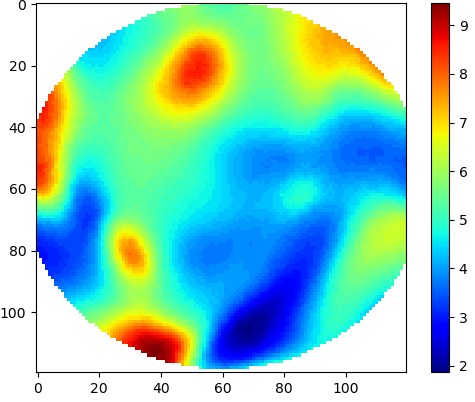}
    \caption{Example of map of characteristic time $\tau_0$ (in seconds), here for a temperature difference $\Delta T=+16^\circ$ with respect to the ambiant temperature.}
    \label{fig:tau0map}
\end{figure}
To understand the time-dependant behaviour of the turbulence, we performed an analysis on each observation (i.e. each series of 512 images for a given temperature).
This analysis consists into defining a characteristic timescale of variation $\tau_0$ for the turbulence in each pixel of the field. We chose to define it arbitrarily as in Equation \ref{eq:deftau}:
\begin{equation}
    \label{eq:deftau}
    \boxed{\frac{R_I(\tau_0)}{R_I(0)}=\frac{1}{2}}
\end{equation}
Where $R_I(\Delta)$ is the auto-correlation of the image matrix $I$ computed for a shift $\Delta$ as defined in Equation \ref{eq:defcor}.
\begin{equation}
\label{eq:defcor}
    R_I(\Delta)=\int I(t).I^*(t-\Delta)~\text{d}t
\end{equation}
We obtain a map of $\tau_0$ over the whole field for a given temperature as shown in Fig.\ref{fig:tau0map}. That is 23 different maps for which we retain the mean value, minimum value and maximum value and display it in Fig.\ref{fig:taus}. It should be noted that the x axis corresponds to the number of the observation. Then the upper figure illustrates the variations of $\tau_0$ in a logarithmic scale while the lower figure shows the temperature  associated to each observation.

\begin{figure} [ht]
   \begin{center}
   \begin{tabular}{c} 
   \includegraphics[width=0.65\linewidth]{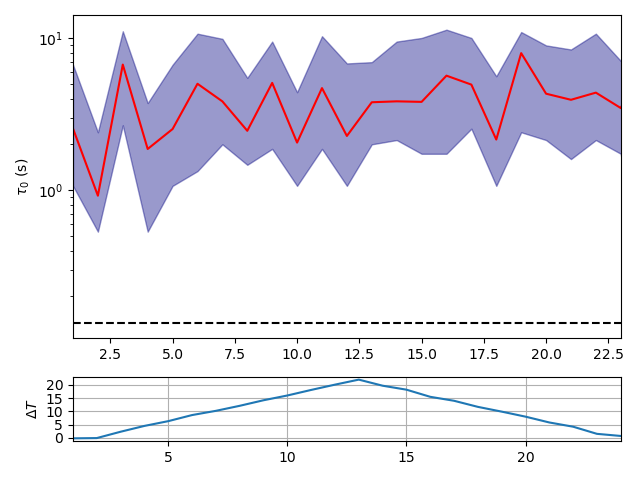}
	\end{tabular}
	\end{center}
   \caption[\label{fig:taus}]
   {\textbf{Upper pannel :} Evolution of $\tau_0$. Red line is the mean value, surrounded by maximum and minimum. Dashed black line shows the acquisition rate (x axis represents the number of the measurement).\\
   \textbf{Lower pannel :} Evolution of temperature for every measurement (x axis represents the number of the measurement).}
   \label{fig:taus}
   \end{figure} 

One can see that the order of magnitude of $\tau_0$ ($\sim 1/10$ s) is well within range of any AO system. But more surprisingly, there is no correlation whatsoever between the timescale of variation of the turbulence and the temperature of the mirror. This result is surprising since one would expect the turbulence to be faster when the optical surface (and the air below it) is heated-up, hence the timescale should decrease when the temperature increases and vice-versa.

\subsection{Space-dependant analysis}
\label{sec:spaceanalysis}
Not only should the turbulence be slow enough so that our AO can follow it, but it should be spatially manageable too.

To make sure that this is the case, we projected our phase maps on the Zernike polynomial basis (limiting it to the 20 firsts polynomials in the Noll convention system). We obtain 22 observations of 512 snapshots for each temperature and for each polynomial. We averaged each time series, so that we get one average observation per temperature per polynomial. For further analysis, we will call "low order" polynomials those bellow the 5$^\text{th}$ Noll Zernike. "intermediate order" polynomials will be higher than the 5$^\text{th}$ and lower than the 10$^\text{th}$ Noll. Finally, "high order" modes will encompass 10$^\text{th}$ Noll and above.

In Fig.~\ref{fig:Zs} we show the evolution of the total power in all the Zernikes. Once again, there is no correlation with the temperature of the optical surface. However, one may expect the power to transfer from the low order to the high order modes when increasing the temperature and the other way around when decreasing it. This only seems to fit with the model of a bubble of air being heated-up, and forcing energy to be dissipated at smaller scales. To check this, we normalized the power in all the Zernikes to get figure \ref{fig:ZNs}. However, no variation whatsoever is observed in the relative power of the different orders. It seems that the heating of the mirror did not trigger more turbulence, nor did it influence the length scale of the pre-existing turbulence.

\begin{figure} [ht]
   \begin{center}
   \begin{tabular}{c} 
   \includegraphics[width=0.65\linewidth]{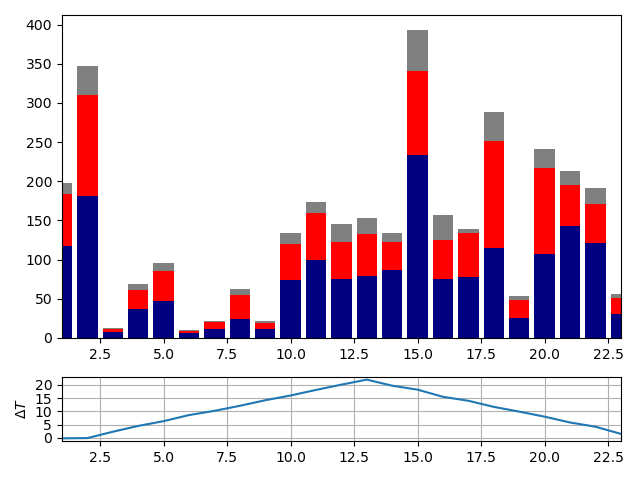}
	\end{tabular}
	\end{center}
   \caption[\label{fig:Zs}]
   {\textbf{Upper pannel :} Evolution of the total power in all the observations projected in the Zernike basis. Blue is for the low order modes, red is for intermediate order modes, and grey is for the high order modes (x axis represents the number of the measurement). The power is expressed in nm$^2$. Typical wavefront residual when including all modes is about 15\,nm.\\
   \textbf{Lower pannel :} Evolution of temperature for every measurement (x axis represents the number of the measurement).}
   \label{fig:Zs}
   \end{figure}
\begin{figure} [ht]
   \begin{center}
   \begin{tabular}{c} 
   \includegraphics[width=0.65\linewidth]{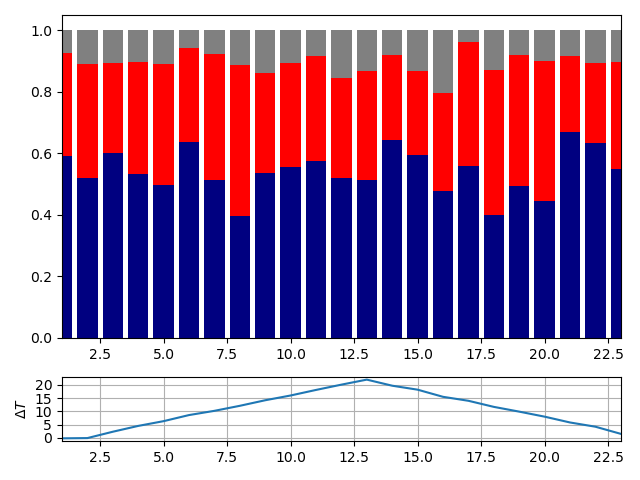}
	\end{tabular}
	\end{center}
   \caption[\label{fig:ZNs}]
   {\textbf{Upper pannel :} Evolution of the normalized power in all the observations projected in the Zernike basis. Blue is for the low order modes, red is for intermediate order modes, and grey is for the high order modes (x axis represents the number of the measurement).\\
   \textbf{Lower pannel :} Evolution of temperature for every measurement (x axis represents the number of the measurement).}
   \label{fig:ZNs}
   \end{figure}

\section{CONCLUSION}
\label{sec:conclusion} 

Both our time-dependant analysis and our spatial analysis led to the conclusion that no influence of the temperature of the optical surface can be measured for our 100mm-diameter mirror, facing down, and mounted in a gimbal mount. We have explored temperature difference with respect to the ambient up to up $+20^\circ$.

To date, we explain this counter intuitive result with the hypothesis that the power injected in the mirror with the heater is mainly evacuated on the upper face of the mirror (see Fig.\ref{fig:monture}), and that the gimbal structure is sufficient to separate the upper flow from the flow in front of the optical surface. Further analysis replacing the flat mirror with an actual Deformable Mirror should confirm these results.

Our experiment validates \emph{a-posteriori} the choice of the NAOMI project to warm up the Deformable Mirror in order to maintain the temperature of the Deformable Mirror around $20^\circ$C (in order to ensure that the device delivers the required performances). We confirm by experiment in laboratory that the additional turbulence generated by the surface temperature being above ambient temperature (up to $+15^\circ$C on the coldest night at Paranal observatory) is negligible.

As a next step, we plan to perform a similar experiment but using the 41x41 Deformable Mirror of GRAVITY+. Instead of heating the mirror with an external heater, we will play typical sequence of correction for bad seeing which may warm up the optical surface and the housing of the Deformable Mirror. The exact value of this warm-up will depend on the actual heat extraction efficiency. The results presented in this work on a reference flat mirror will be an interesting reference point to compare with the the measurement obtained with the real Deformable Mirror.

In a nutshell, we performed an experiment which demonstrates that heating a facing-down mirror up to $+20^\circ$ above ambient temperature does not trigger significant turbulence. Hence the possibility to discuss the general statement that every optical surface should be kept within $\pm$1.5$^\circ$ around ambient temperature.

\acknowledgments 
 
We address our greatest thanks to the Laboratoire Lagrange (Nice, France) and its members for their implication in this work.\\
This work is supported by the Action Spécifique Haute Résolution Angulaire (ASHRA) of CNRS/INSU co-funded by CNES.\\
This work is supported by the French National Research Agency in the framework of the “Investissements d’avenir” program (ANR-15-IDEX-02)\\
Finally, we thank the SPIE organisation for permitting us to present this work to the rest of the community, which allowed new fruitful development prospects.
\bibliography{report} 
\bibliographystyle{spiebib} 

\end{document}